# Band alignment of grafted monocrystalline Si (001)/$\beta$-Ga$_2$O$_3$ (010) p-n heterojunction determined by X-ray photoelectron spectroscopy

Jiarui Gong[1,a)], Jie Zhou[1,a)], Ashok Dheenan[2], Moheb Sheikhi[1], Fikadu Alema[3], Tien Khee Ng[4], Shubhra S. Pasayat[1], Qiaoqiang Gan[4], Andrei Osinsky[3], Vincent Gambin[5], Chirag Gupta[1], Siddharth Rajan[2,b)], Boon S. Ooi[4,b)], and Zhenqiang Ma[1,b)]

[1]*Department of Electrical and Computer Engineering, University of Wisconsin-Madison, Madison, Wisconsin, 53706, USA*

[2]*Department of Electrical and Computer Engineering, The Ohio State University, Columbus, OH 43210, USA*

[3]*Agnitron Technology Incorporated, Chanhassen, MN 55317, USA*

[4]*Department of Electrical and Computer Engineering, King Abdullah University of Science and Technology, Thuwal 23955-6900, Saudi Arabia*

[5]*Northrop Grumman Corporation, Redondo Beach, CA 90278, USA*

a) Jiarui Gong and Jie Zhou contributed equally to this work.
b) Author to whom correspondence should be addressed. Electronic mail: mazq@engr.wisc.edu, boon.ooi@kaust.edu.sa, rajan.21@osu.edu

**Abstract**

Beta-phase gallium oxide ($\beta$-Ga$_2$O$_3$) research has gained accelerated pace due to its superiorly large bandgap and commercial availability of large-diameter native substrates. However, the high acceptor activation energy obstructs the development of homojunction bipolar devices employing $\beta$-Ga$_2$O$_3$. The recently demonstrated semiconductor grafting technique provides an alternative and viable approach towards lattice-mismatched $\beta$-Ga$_2$O$_3$-based p-n heterojunctions with high quality interfaces. Understanding and quantitatively characterizing the band alignment of the grafted heterojunctions is crucial for future bipolar device development employing the grafting method. In this work, we present a systematic study of the band alignment in the grafted monocrystalline Si/$\beta$-Ga$_2$O$_3$ heterostructure by employing X-ray photoelectron spectroscopy (XPS). The core level peaks and valence band spectra of the Si, $\beta$-Ga$_2$O$_3$, and the grafted heterojunction were carefully obtained and analyzed. The band diagrams of the Si/$\beta$-Ga$_2$O$_3$ heterostructure were constructed using two individual methods, the core level peak method and the valence band spectrum method, by utilizing the different portions of the measured data. The reconstructed band alignments of the Si/$\beta$-Ga$_2$O$_3$ heterostructure using the two different methods are identical within the error range. The band alignment is also consistent with the prediction from the electron affinity values of Si and $\beta$-Ga$_2$O$_3$. The study suggests that the interface defect density in grafted Si/$\beta$-Ga$_2$O$_3$ heterostructure is at a sufficiently low level such that Fermi level pinning at the interface has been completely avoided and the universal electron affinity rule can be safely employed to construct the band diagrams of grafted monocrystalline Si/$\beta$-Ga$_2$O$_3$ heterostructures.

**Introduction**

The fourth generation semiconductors, also known as the ultrawide-bandgap semiconductors, have tremendous potential for future power electronics, extreme-environment electronics, deep-ultra-violet optoelectronics, etc., originating from their superiorly large bandgap energy among other outstanding electronical and thermal properties [1]. Significant research progress has been made in the beta-phase gallium oxide ($\beta$-$Ga_2O_3$) in recent years, partly due to the commercial availability of large diameter native substrates [2-4]. Unipolar devices with outstanding performances, such as Schottky barrier diodes [5-9] and field effect transistors [10, 11], have been demonstrated and developed rapidly.

However, the extraordinarily high ionization energy of acceptors [12-14] prevents achieving efficient p-type doping and thus the development of homojunction bipolar devices employing $\beta$-$Ga_2O_3$ has been impeded. To bypass the p-type doping challenge, several approaches have been demonstrated towards achieving heterojunctions by combining other p-type materials with the n-type $\beta$-$Ga_2O_3$, such as p-n junctions of $NiO_x$/$Ga_2O_3$ [15, 16], Si/$Ga_2O_3$ [17-19], and GaAs/$Ga_2O_3$ [20], etc. Among these approaches, semiconductor grafting technique [21] can enable high-quality interface between a monocrystalline p-type semiconductor with the n-type $\beta$-$Ga_2O_3$, such as Si/$Ga_2O_3$ [17, 19] and GaAs/$Ga_2O_3$ [20]. The high interface quality is manifested in the low p-n diode ideality factor and low interface trap density.

To extend the grafting technique into advanced bipolar device applications employing the grafted p-n heterojunctions building blocks between monocrystalline p-type semiconductors and n-type $\beta$-$Ga_2O_3$, the band alignment of the heterojunctions ought to be characterized and fully understood. While the grafted interfaces [17, 19, 20] have exhibited sufficiently low density of interface traps ($D_{it}$), an interface with a high $D_{it}$ value can cause Fermi level pinning at the interface and lead to deviation of its band alignment from a theoretical prediction. In other words, a characterized ideal band alignment can also prove the existence of a high-quality interface with a sufficiently low $D_{it}$. In our previous work [19], we have conducted electrical current-voltage and capacitance-voltage measurements on the grafted Si/Ga$_2$O$_3$ p-n heterojunction to study the charge carrier properties around the interface. The ideal rectifying behaviors and the frequency-independent capacitance-voltage characteristics both suggested a high quality heterointerface with sufficiently low $D_{it}$. The band diagram was constructed based on the reported electron affinity values of Si and Ga$_2$O$_3$ but lacks experimental proof for the construction. A direct measurement of the band alignment between Si and Ga$_2$O$_3$ would vouch whether the universal electron affinity rule is applicable to the grafted monocrystalline Si/Ga$_2$O$_3$ heterostructures. An experimentally proven applicability of the electron affinity rule to the grafted Si/Ga$_2$O$_3$ heterojunction would generate a broad impact in the Ga$_2$O$_3$ device community. Therefore, a systematic study of the band offsets between the grafted monocrystalline Si and Ga$_2$O$_3$ becomes highly desirable.

In this work, we employed X-ray photoelectron spectroscopy (XPS) to investigate the band alignment in the grafted monocrystalline Si/$\beta$-Ga$_2$O$_3$ heterojunction. The core level peaks and valence band spectra of Si, $\beta$-Ga$_2$O$_3$ and Si/$\beta$-Ga$_2$O$_3$ heterostructures were collected from XPS. After calculating the bandgap of $\beta$-Ga$_2$O$_3$ from the XPS data, two different methods were adopted to calculate the valence band offset (VBO) and conduction band offset (CBO) between Si and $\beta$-Ga$_2$O$_3$ to cross-check the reliability of the results. The band diagrams were separately constructed using the two methods, which demonstrated excellent consistency between each other. Of more importance, the band offset values in the grafted Si/$\beta$-Ga$_2$O$_3$ heterostructure match that expected from the electron affinity values of Si and $\beta$-Ga$_2$O$_3$. Therefore, based on these observations, it can be posited that using the electron affinity rule for band alignment construction is applicable to the grafted monocrystalline Si/*β*-Ga₂O₃ heterostructure.

**Experiment**

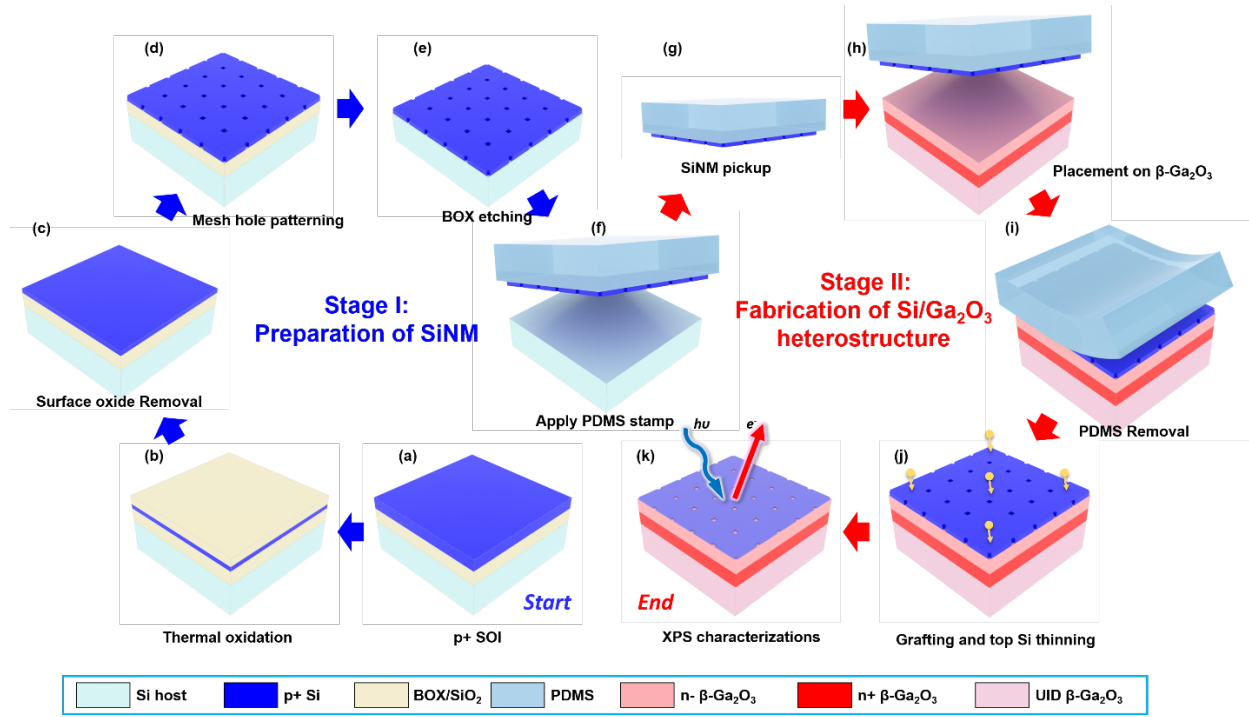

**Fig. 1.** Schematic illustration of the fabrication procedure of the grafted monocrystalline Si/Ga$_2$O$_3$ heterostructure sample for the X-ray photoelectron spectroscopy (XPS) characterization. (a) Initial silicon-on-insulator (SOI) structure, consisting of a 180 nm thick top p$^+$ Si layer, a 400 nm buried oxide layer (BOX), and a Si handling substrate. (b) Thermal oxidation of the p$^+$ SOI wafer to oxidize ~130 nm top Si into ~300 nm thermal SiO$_2$ on its top. (c) Removal of the surface SiO$_2$, exposing the underlying 50 nm p$^+$ Si layer. (d) Creation of meshed etching holes in the top p$^+$ Si layer. (e) Removal of the BOX layer using hydrofluoric acid. (f) Pick up the p+ Si NM from the host substrate via a polydimethylsiloxane (PDMS) stamp. (g)-(i) Transfer of the released SiNM to the Ga$_2$O$_3$ substrate, followed by rapid thermal annealing process at 350 °C for 5 mins. (j)

Reduction of the SiNM thickness down to ~10 nm through slow RIE dry etching. (k) The finalized grafted monocrystalline Si/Ga$_2$O$_3$ heterostructure was subjected to XPS characterizations.

The regular fabrication process of the Si/β-Ga$_2$O$_3$ p-n heterostructure that consists of a 180 nm top p$^+$-Si layer has been recorded elsewhere [19]. However, because the sampling depth of XPS is only ~10 nm below the sample surface, additional steps are required to validate the sample for XPS measurements. Conventionally, the preparation of such samples can follow bottom-up approaches, such as employing epitaxy method or deposition method [22-24] to form the 10 nm surface layer. But in this work, to analyze the interface of the grafted Si/β-Ga$_2$O$_3$ heterostructure by XPS, only a top-down approach is viable, *i.e.*, the top Si layer should be thinned down from its original thickness of 180 nm down to 10 nm while maintaining its single crystallinity. To fully illustrate the Si/β-Ga$_2$O$_3$ heterostructure sample preparation for the XPS study, Fig. 1 depicts the entire fabrication process flow. It should be mentioned that except for the top Si NM thinning down process, all other fabrication procedures and process parameters/conditions are identical to those in our previous report [19]. Therefore, it is expected that the interface quality of this heterostructure sample is equivalent to that of the previously characterized one [19].

The sample fabrication process can be divided into two major stages: the preparation of Si nanomembrane (Si NM) and the formation Si/Ga$_2$O$_3$ heterostructure. The first stage begins with a silicon-on-insulator (SOI) sample, as shown in Fig. 1(a). This starting SOI consists of a top 180 nm thick heavily doped Si layer (B doped, $5\times10^{19}$ cm$^{-3}$), a 400 nm buried oxide (BOX) layer, and a Si handling substrate. To perform the thermal oxidation process for thinning the Si layer, the sample was first cleaned using standard RCA cleaning procedure [19] to remove surface contaminants and native oxide, thereby ensuring a pristine surface. The as-cleaned SOI was then loaded into the MRL High Temp Oxidation Furnace and subjected to thermal oxidation at 1050 °C for 6.5 hours, resulting in a ~300 nm thermal oxide layer by consuming 130 nm thick Si, leaving a remaining 50 nm thick $p^+$ Si, as shown in Fig. 1(b). This thermal oxide layer was removed using 49% hydrofluoric acid (HF) for 3 minutes, exposing the 50 nm $p^+$ Si (named as 50 nm SOI sample), as shown in Fig. 1(c). The SOI was subsequently patterned into an array of $9 \times 9$ μm$^2$ square holes spaced by 55 μm gaps, using standard photolithography and reactive ion etching (RIE) (Fig. 1(d)). This etching exposed the 400 nm BOX layer, preparing it for the subsequent chemical undercut. Next, the BOX layer was chemically etched away using 49% HF solution for 2 hours, producing a 50 nm thick Si NM sitting on the Si handling substrate, as shown in Fig. 1(e).

In the second stage of grafting Si with $Ga_2O_3$, the Si NM was delicately retrieved from its Si host using a soft polydimethylsiloxane (PDMS) stamp, as shown in Fig. 1(f). The Si NM in Fig. 1(g) was then transfer-printed to the epitaxial $Ga_2O_3$ destination substrate, as shown in Fig. 1(h). The detailed growth [25], specification [25], and surface treatment of the epitaxial $Ga_2O_3$ substrate [19] are detailed in our previous works. It should be highlighted that in our heterostructure formation process, the surface oxygen composition, as the key element for passivation, was enhanced as verified by XPS characterizations [19]. This is an important step towards the later formation of the interfacial oxide layer ($SiGaO_x$ here) that has provided sufficient simultaneous (or double-side) passivation for both Si and $β$-$Ga_2O_3$, ensuring the exceptional quality of Si/$β$-$Ga_2O_3$ interface. After removing the PDMS stamp (Fig. 1(i)), a rapid thermal annealing (RTA) process was conducted at 350 °C for 5 minutes to form a robust chemical bonding between the SiNM and $Ga_2O_3$ substrate. Note that the thermal annealing condition is another key process control parameter that led to the high Si/$β$-$Ga_2O_3$ interface quality. Different thermal budgets applied to the Si/$β$-$Ga_2O_3$ heterostructure to form chemical bonds in between the two materials may result in very different $D_{it}$ values at the interface, thereby creating different band alignment between Si and *β*-Ga$_2$O$_3$, deviating from the results presented in this study. The thermal budget selected in our Si/*β*-Ga$_2$O$_3$ heterostructure studies was based on our prior grafting experiments [21]. The Si NM was then further thinned from 50 nm down to ~10 nm by a precisely controlled slow dry etching process using a PlasmaTherm 790 RIE etcher (Fig. 1(j)). A completed Si/Ga$_2$O$_3$ heterostructure (named as the grafted Si/Ga$_2$O$_3$ sample) was then ready for the subsequent XPS characterization, as shown in Fig. 1(k).

To construct the Si/*β*-Ga$_2$O$_3$ band diagram, the XPS measurements were performed on the pristine Ga$_2$O$_3$ substrate, the 50 nm SOI sample (Fig. 1(c)), and the grafted Si/Ga$_2$O$_3$ heterostructure sample (Fig. 1(k)) through a Thermo Scientific K Alpha X-ray Photoelectron Spectrometer (XPS) with an Al K$_α$ X-ray source ($hv$ = 1486.6 eV). The instrument was calibrated by Au 4$f_{7/2}$ peak at 84.0 eV. The following settings were applied to the spectrometer: 10 eV pass energy, 400 μm spot size, up to 10 s dwell time, and 0.02 eV step size. Details regarding sample handling can be found in our previous work [19].

**Results and discussion**

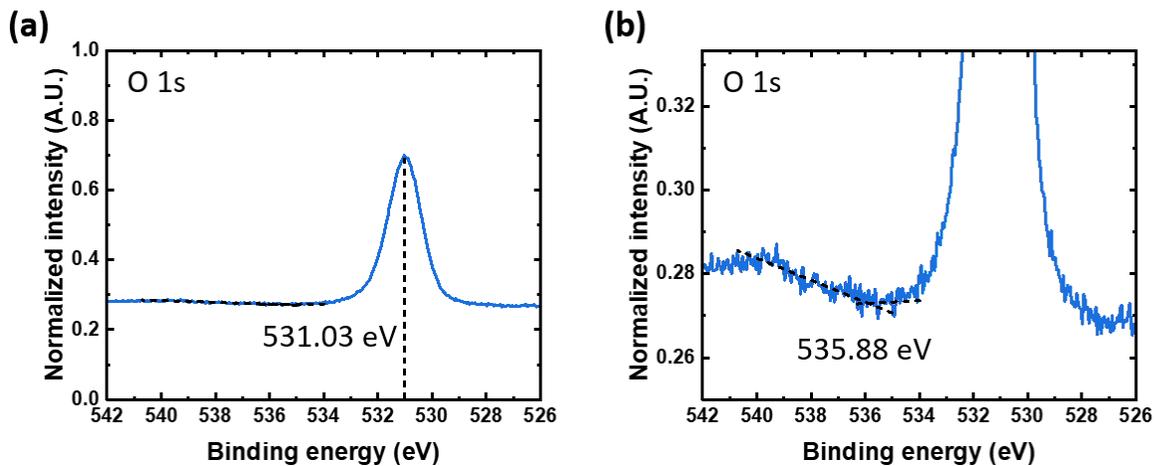

**Fig. 2.** (a) The O 1$s$ XPS spectrum obtained from the pristine Ga$_2$O$_3$ substrate with extraction of the O 1$s$ peak position. (b) The enlarged O 1$s$ XPS spectrum of the pristine Ga$_2$O$_3$ substrate with extraction of the onset of inelastic energy loss.

To first measure the bandgap value of the Ga$_2$O$_3$, the O 1$s$ XPS spectrum was obtained from the pristine Ga$_2$O$_3$ substrate. As shown in Fig. 2(a), the O 1$s$ peak is centered at a binding energy of 531.03 eV. The onset of inelastic energy loss is extracted at 535.88 eV through linear regression (Fig. 2(b)). Thus, the bandgap of Ga$_2$O$_3$ can be estimated to be 4.85 eV, i.e., the binding energy difference between the core level peak (531.03 eV) and the onset of inelastic energy loss (535.88 eV). This bandgap energy value is consistent with the previously reported value by optical transmission measurements [26], confirming the reliability of the XPS measurement results.

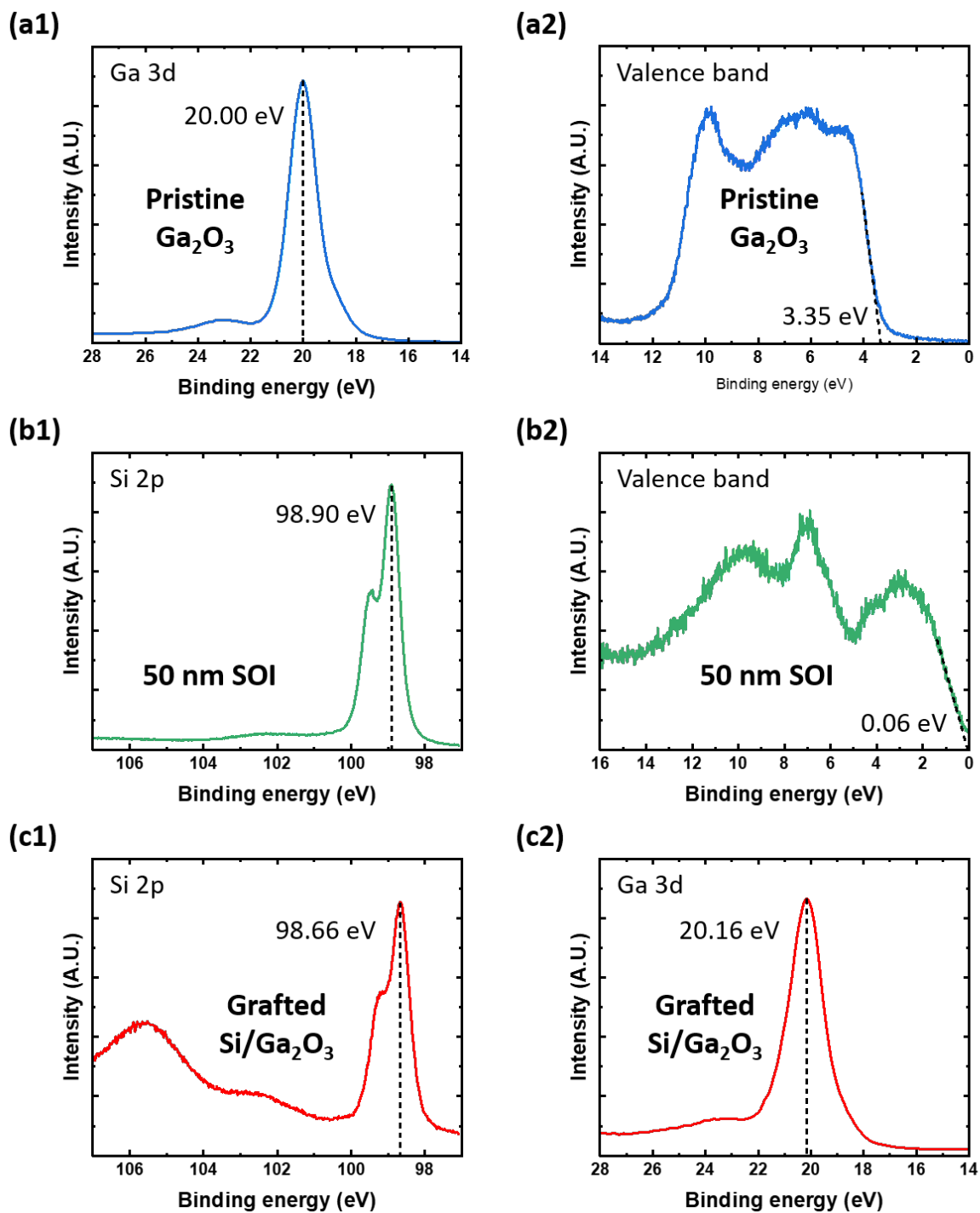

**Fig. 3.** (a1) The Ga 3*d* XPS spectrum collected from the pristine $Ga_2O_3$ substrate with extraction of the Ga 3*d* peak position. (a2) The valence band XPS spectrum of the pristine $Ga_2O_3$ substrate with extraction of the valence band maximum (VBM). (b1) The Si 2*p* XPS spectrum obtained

from the 50 nm SOI sample with extraction of the Si $2p_{3/2}$ peak position. (b2) The valence band XPS spectrum of the 50 nm SOI sample with extraction of the VBM. (c1) The Si $2p$ XPS spectrum of the grafted Si/Ga$_2$O$_3$ sample with extraction of the Si $2p_{3/2}$ peak position. (c2) The Ga $3d$ XPS spectrum of the grafted Si/Ga$_2$O$_3$ heterostructure sample with extraction of the Ga $3d$ peak position.

The valence band offset between Si and the $\beta$-Ga$_2$O$_3$ can be measured through the method developed by Kraut *et al.* in 1980 [27]. From the pristine Ga$_2$O$_3$ substrate, the Ga $3d$ core level peak was measured at 20.00 eV (Fig. 3(a1)) and the valence band maximum is determined at 3.35 eV (Fig. 3(a2)). From the 50 nm thick SOI sample, the Si $2p_{3/2}$ core level peak was measured at 98.90 eV (Fig. 3(b1)) and the VBM was determined at 0.06 eV. From the grafted Si/$\beta$-Ga$_2$O$_3$ heterostructure, the Si $2p_{3/2}$ core level peak was measured at 98.66 eV and the Ga $3d$ core level peak appeared at 20.16 eV. Therefore, following the equation proposed by Kraut *et al.*[27]

$$\Delta E_V = (E_{CL}^Y - E_V^Y) - (E_{CL}^X - E_V^X) - \Delta E_{CL}, \quad (1)$$

where $\Delta E_V = E_V^X - E_V^Y$ is the valence band offset, $\Delta E_{CL} = E_{CL}^Y(i) - E_{CL}^X(i)$ is the binding energy difference between core level (CL) peaks of semiconductor X and Y at the interface, $E_{CL}^Y - E_V^Y$ is the binding energy difference between the core level peak and VBM in semiconductor Y, and $E_{CL}^X - E_V^X$ is the binding energy difference between the core level peak and VBM in semiconductor X. In this study, Ga$_2$O$_3$ represents semiconductor X and Si the semiconductor Y. The corresponding VBO value was calculated to be 3.63 eV. Combining this result with the measured Ga$_2$O$_3$ bandgap value (4.85 eV) and the reported Si bandgap value (1.12 eV [28]), the CBO value between Si and $\beta$-Ga$_2$O$_3$ was calculated to be -0.10 eV.

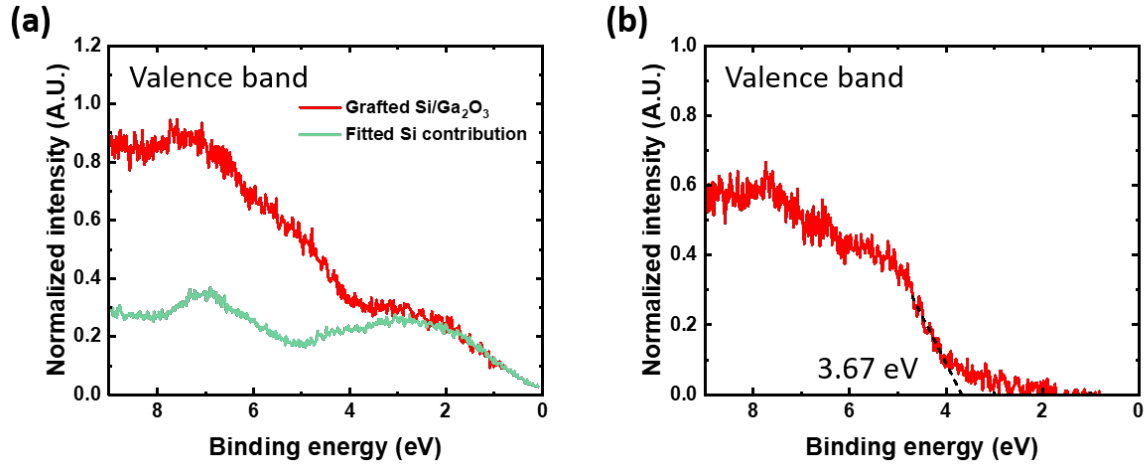

**Fig. 4.** (a) The valence band XPS spectrum of the grafted Si/Ga$_2$O$_3$ sample (red curve). The light red curve shows the fitted spectrum contribution from the Si valence band. (b) The valence band XPS spectrum of the grafted Si/$\beta$-Ga$_2$O$_3$ heterostructure sample after subtraction of the fitted spectrum contribution from Si valence band, with the extraction of the VBM.

To cross verify the band offset values obtained using the Kraut *et al.* method, we adopted an alternative method that solely depends on the valence band spectra [29] (Fig. 4). The red curve in Fig. 4(a) shows the valence band XPS spectrum of the grafted Si/Ga$_2$O$_3$ sample. The spectrum is a mixture of signals from both Si and Ga$_2$O$_3$. The spectrum contribution from Si (light red curve in Fig. 4(a)) was fitted by multiplying a fitting factor with the Si valence band spectrum in Fig. 3(b2). The Ga$_2$O$_3$ valence band component was therefore obtained after subtracting the fitted Si spectrum component (Fig. 4(b)), demonstrating a VBM value of 3.67 eV. Therefore, the VBO value can be directly calculated from the Si VBM (0.06 eV) and Ga$_2$O$_3$ VBM (3.67 eV) as 3.61 eV. The corresponding CBO value between the Si and the $\beta$-Ga$_2$O$_3$ was calculated as -0.12 eV.

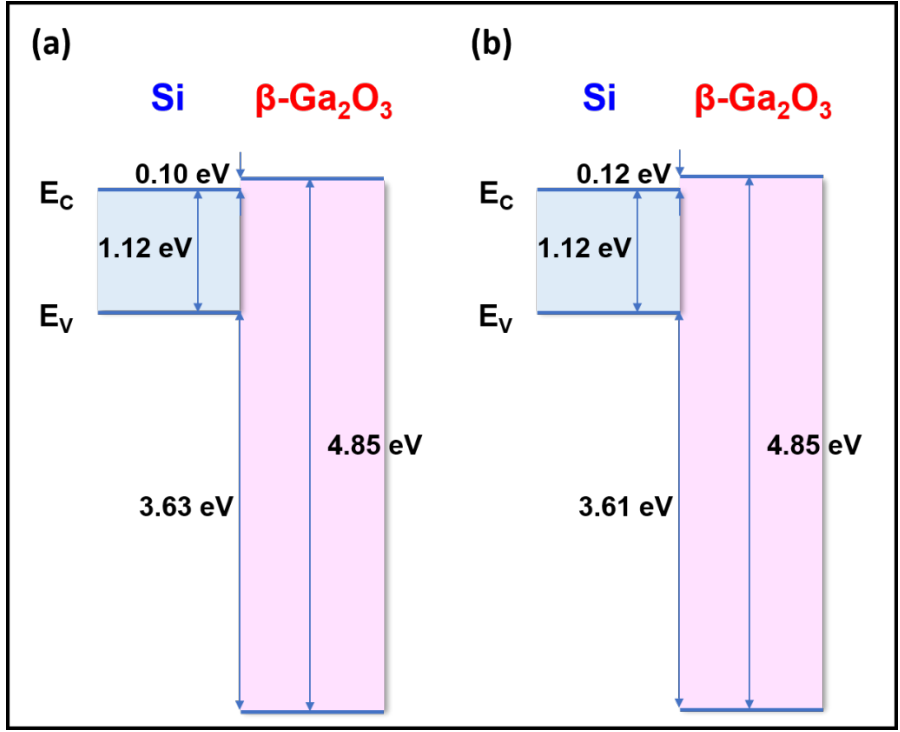

**Fig. 5.** The band diagram constructed for the grafted monocrystalline Si/$β$-Ga$_2$O$_3$ heterostructure from methods of (a) core level peaks [27] and (b) valence band spectra [29]. The two methods show identical band alignment within the measurement error range. The band alignment between Si and $β$-Ga$_2$O$_3$ in the grafted Si/$β$-Ga$_2$O$_3$ heterostructure concord with the electron affinity rule, indicating that the grafted interface is free of Fermi level pinning.

The band diagrams of the grafted monocrystalline Si/$β$-Ga$_2$O$_3$ heterostructure based on the results obtained from the above two methods were reconstructed accordingly. As shown in Fig. 5, the band offset values and the band alignment constructed using the two methods, which utilize different sets of XPS data (Fig. 3 and Fig. 4), are essentially identical within the measurement errors. Of more importance, the results are also consistent (also within the error range) with the theoretical CBO value that can be predicted by applying the electron affinity rule to the individual Si (4.05 eV [28]) and $β$-Ga$_2$O$_3$ (4.00 eV [30]). The experimental measurements clearly imply that $D_{it}$ has been made sufficiently low in the grafted Si/$β$-Ga$_2$O$_3$ heterostructure, thereby Fermi level pinning at the interface has been circumvented. This epitaxy quality interface like the ones formed in lattice-matched epitaxial interface enables simple, electron affinity rule-based band alignment construction.

**Conclusion**

In this work, a systematic study of the band alignment in the monocrystalline grafted Si/$β$-Ga$_2$O$_3$ heterostructure through XPS measurements was conducted. The bandgap of $β$-Ga$_2$O$_3$ was determined to be 4.85 eV. The VBO and CBO values between Si and $β$-Ga$_2$O$_3$ were calculated from the method of core level peaks and the method of valence band spectrum, which resulted in identical band alignment and complied with the electron affinity rule. The experiment suggests that the grafted Si/$β$-Ga$_2$O$_3$ interface is free of $D_{it}$ induced Fermi leveling pining effects, and thus the interface is of high quality. The work may inspire further device development using the grafting approach and will benefit the future development of $β$-Ga$_2$O$_3$-based devices.

**Acknowledgements**

The work was supported by a CRG grant (2022-CRG11-5079.2) by the King Abdullah University of Science and Technology (KAUST) and by Air Force Office of Scientific Research under grant FA9550-21-1-0081. It was partially supported by the Office of Naval Research (ONR, Program Manager: Mr. Lynn Petersen) and AFWERX through contract numbers N6833518C0192 and FA864921P0304, respectively.

**Data availability statement**

The data that support the findings of this study are available from the corresponding author upon reasonable request.